\documentclass[acmsmall]{acmart}
\AtBeginDocument{%
  \providecommand\BibTeX{{%
    \normalfont B\kern-0.5em{\scshape i\kern-0.25em b}\kern-0.8em\TeX}}}

\usepackage{hhline}

\setcopyright{acmcopyright}
\copyrightyear{2023}
\acmYear{2023}
\acmDOI{XXXXXXX.XXXXXXX}

\begin{document}

\title{Disinformation 2.0 in the Age of AI: A Cybersecurity Perspective}



\author{Wojciech Mazurczyk}
\email{wojciech.mazurczyk@pw.edu.pl}
\affiliation{%
  \institution{Warsaw University of Technology}
  \streetaddress{Nowowiejska 15/19}
  \city{Warsaw}
  \state{Mazovia}
  \country{Poland}
  \postcode{00-665}
}

\author{Dongwon Lee}
\email{dongwon@psu.edu}
\affiliation{%
  \institution{The Pennsylvania State University}
  \country{USA}
}

\author{Andreas Vlachos}
\email{av308@cam.ac.uk}
\affiliation{%
  \institution{University of Cambridge}
  \city{Cambridge}
  \country{UK}
}

\renewcommand{\shortauthors}{Mazurczyk et al.}

%



\maketitle

\section{Disinformation and Its Evolution}
According to a report from Lloyd's Register Foundation\footnote{
The Lloyd's Register Foundation World Risk Poll (2019), 
https://tinyurl.com/znjya6ct}, at present, cybercrime is one of the biggest concerns of Internet users worldwide, with disinformation\footnote{In this work, among related concepts and terms, we use the term, \emph{disinformation}, to refer to ``false information created with malicious intention," per \cite{Kim2021}.
} ranking highest among such risks (57\% of internet users across all parts of the world, socio-economic groups, and all ages). 
Yet, for years, there has been a discussion in the security community about whether disinformation should be considered a cyber threat \cite{Zurko2022}. 
However, recent worldwide phenomena (e.g., the increase in the frequency and sophistication of cyberattacks, the 2016 US election interference, and Russian invasion in Ukraine, the COVID-19 pandemic, etc.) have made disinformation one of the most potent cybersecurity threats for businesses, governments, the media, and whole society as a whole. 
In addition, recent breakthroughs in AI have further enabled the creation of highly realistic fake contents at scale.
As such, we argue that disinformation should be rightfully considered a cyber threat, and therefore developing effective countermeasures is critically necessary.

The way that disinformation is evolving is not exactly new, as other ``classical'' cyber threats followed a similar path. First, disinformation has been around for centuries, and the internet is just the latest means of communication used to spread it. We have already witnessed similar developments before. That is, there were different types of crimes like scams, extortions, and thefts, 
and we now see their cyber versions, which are less risky for attackers yet more effective than their classical forms.  Thus, the use of the internet (especially  social media) made it possible to boost the scale and range at which these attacks can be launched, but the essence of the attack itself remains the same. Similarly, disinformation
can affect many more people in a much shorter time than in the case of the non-digital version (e.g., traditional newspapers, TV news).
Moreover, advances in AI allowed the creation of deepfakes in various types of digital media (i.e., images, video, speech) and text, and the introduced modifications are tough to spot, distinguish, and explain 
This greatly enhances the resulting disinformation's potential reach and believability. 
%

Note that disinformation is not necessarily expected to provide direct revenue, as in the case of other cyber threats. However, such cases already have happened, e.g., by spreading disinformation to manipulate stock price\footnote{NBCnews, ``SEC Cracks Down on Fake Stock News'', 2017, URL: https://www.nbcnews.com/business/markets/sec-cracks-down-fake-stock-news-n745141} or earning income by disseminating it\footnote{Heather C. Hughes, Israel Waismel-Manor, 'The Macedonian Fake News Industry and the 2016 US Election', URL: https://www.cambridge.org/core/journals/ps-political-science-and-politics/article/macedonian-fake-news-industry-and-the-2016-us-election/79F67A4F23148D230F120A3BD7E3384F}.

\section{Disinformation 2.0}
In the conventional paradigm of disinformation, on the attack side, we have disinformation creators who fabricate false information and post it to websites or social media for various purposes such as monetary incentives or political agenda. 
On the defense side, platforms have used human operators as well as computational methods to ensure the integrity of information, such as disinformation detectors to filter out questionable content, and socialbot detectors to curb the spread of disinformation. By and large, so far, creators (semi-)manually have created and disseminated disinformation content without using sophisticated AI techniques. When fake content is detected and filtered out by defense mechanisms, creators would simply attempt to re-disseminate it using different accounts or domains.

\begin{figure}[tb]
  \centering
  \includegraphics[width=0.8\linewidth]{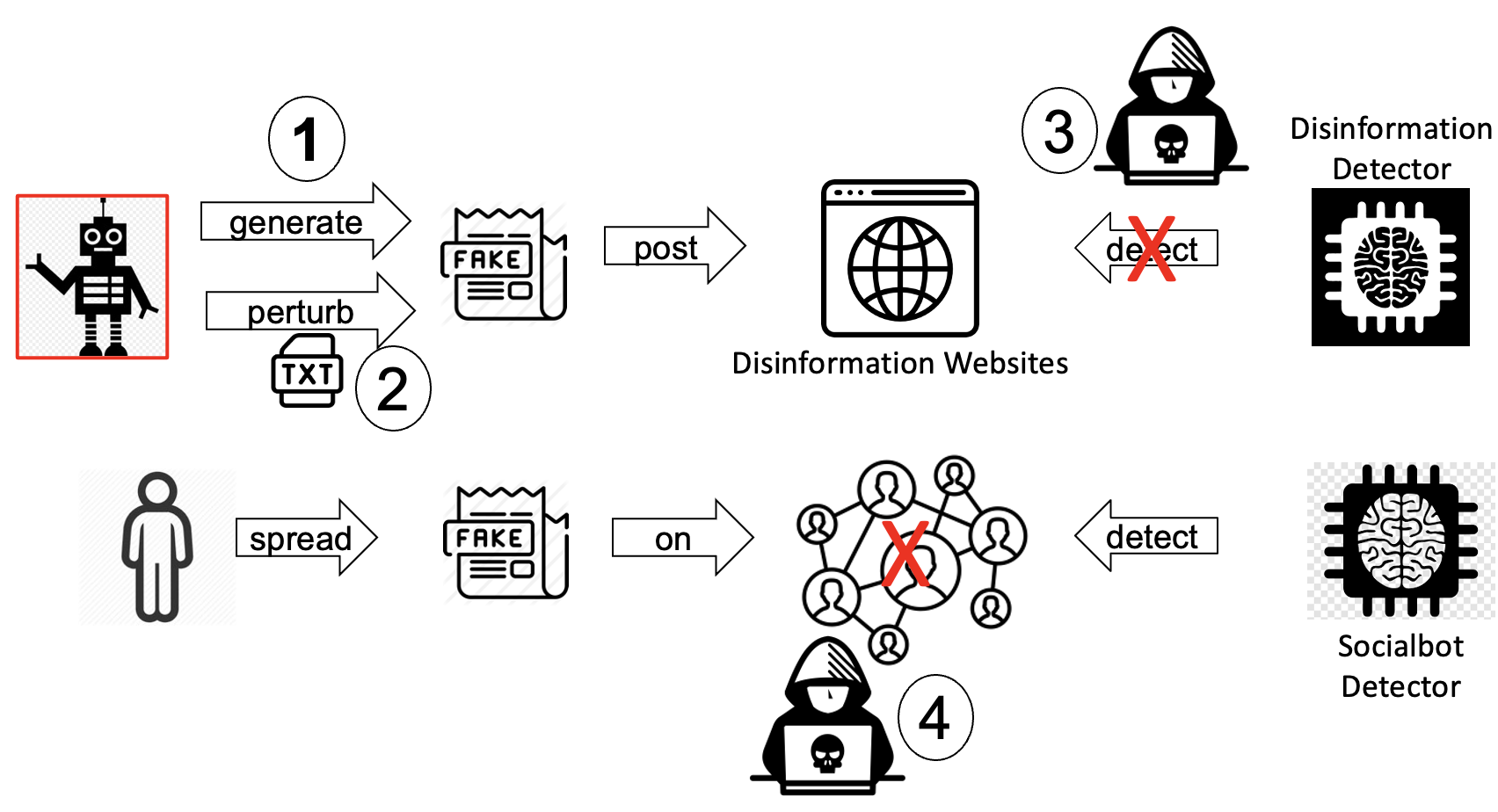}
  \caption{Four plausible attack scenarios under disinformation 2.0.}
  \label{fig:fakescenarios}
\end{figure}

However, with the recent advances in AI, we envision that this paradigm is likely to change substantially, yielding what we call {\bf disinformation 2.0}, where \emph{disinformation would become more targeted and personalized, its content indistinguishable from genuine content, and its creation and dissemination further accelerated by AI}. Disinformation 2.0 will increase distrust in news that humans encounter in real and digital worlds, which is a major problem already\footnote{Digital News Report 2022, URL: https://reutersinstitute.politics.ox.ac.uk/digital-news-report/2022}. Typically, in cyberspace, an attacker's aim is to find weak spots in the targeted system and exploit/compromise them using technical and/or non-technical means. If we treat disinformation as a cyberattack, then several scenarios of disinformation 2.0 become possibilities, as illustrated in Figure~\ref{fig:fakescenarios}: 
\begin{enumerate}
    \item Adversaries generate more convincing disinformation,  using generative AI techniques (e.g., ChatGPT for texts or DALL-E for images),
    \item Adversaries subtly perturb existing content using AI methods to create more convincing disinformation with better capability to evade detection \cite{le2022-per}.
    \item Adversaries use AI methods to attack disinformation detection eco-systems, promoting disinformation and demoting real news \cite{du2022synthetic},
    \item Adversaries strategize the spread of disinformation using AI methods, maximizing its influence on the (social) network while evading socialbot detection eco-systems \cite{Le2022}. 
\end{enumerate}

When the creation and dissemination of disinformation are greatly enhanced using advanced AI methods in one of these scenarios, the resulting disinformation 2.0 becomes much harder to detect and more persuasive/impactful than the previous disinformation 1.0.

\section{Countermeasures}
Detecting disinformation 1.0 has been heavily researched in recent years (e.g., see \cite{Kim2021} for survey), and many solutions have reported high detection accuracies (e.g., \cite{Cui2020}). However, there are still several remaining issues, including early detection, multi-lingual/multi-platform detection, better explainability, or socio-technical issues \cite{survey-AV}. As with every cyber threat, completely eliminating disinformation is unlikely (as achieving complete security is never possible). However, we need to diminish the impact of disinformation on Internet users, as we did with threats like spam emails. Several years ago, for instance, spam emails were considered one of the major threats, but now their scale and relevancy are not as high as they were before \cite{Ferrara2019}. This has been achieved due to decades of research advances, during which many sophisticated techniques enabled to significantly limit the volume of spam emails in Internet users' inboxes. 

Currently, a major disadvantage in our defense against disinformation 2.0 is that dedicated defenses against disinformation are being individually researched, developed, deployed, and evaluated, which is not very effective in diminishing the threat of disinformation. For this challenge, we argue to use
some lessons learned from cybersecurity, where typically multiple ``layers of defense'' are envisioned. On the Internet, as security risks occur at various levels, it is necessary to set up security mechanisms that provide multiple layers of defense against these risks (e.g., by considering threats at the system, network, application, and transmission levels).

A popular approach to layered security is defense-in-depth architecture where controls are designed to protect the physical (i.e., prevent physical access to IT systems, such as security guards or locked doors), technical (i.e., security measures that use specialized hardware or software, e.g., as a firewall, IDS/IPS, antivirus), and administrative (i.e., policies or procedures directed at an organization's employees) aspects of the communication network. 
Using such a layered approach to provide network security makes it possible for an attacker who penetrates one layer of defense to be stopped by a subsequent layer. Therefore, addressing the disinformation problem will also likely require a layered approach where both human-oriented (e.g., raising awareness, training, etc.) and technical measures are applied at the news creation, transmission, and consumption layers.

\begin{figure}[tb]
  \centering
  \includegraphics[width=0.9\linewidth]{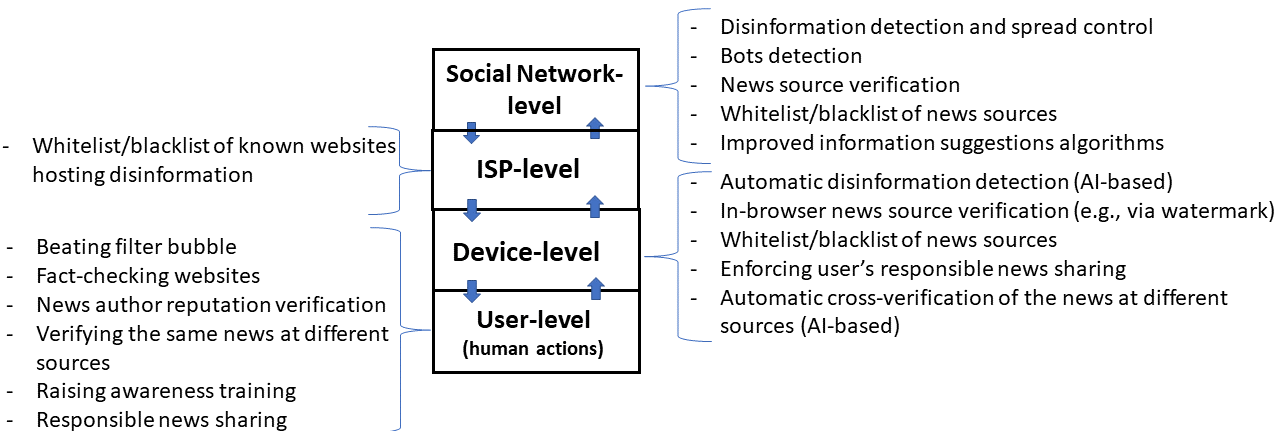}
  \caption{Proposed layered approach to counter disinformation in a holistic manner.}
  \label{fig:fakeproposed}
\end{figure}

Disinformation 2.0 techniques urge to devise new countermeasures. To achieve this, a layered approach could address such a threat in a holistic manner. That is why, we propose to distinguish four layers at which disinformation impact can be diminished (Fig.~\ref{fig:fakeproposed}):
\begin{itemize}
    \item \textit{Social Network-level layer} is organized by the social network operator, who may be equipped with  attack-resistant solutions like (AI-based and/or human operator-aided) disinformation detection, spread control mechanisms, and detection of strategized bots. Additional methods include, for example, verificationtion of news sources 
    (e.g., using reputation-based or digital watermarking-based solutions \cite{megias2022}), whitelisting and blacklisting of news sources, and fixing the information suggestion algorithms to avoid creating filter bubbles. This would be reminiscent of how network topology is taken into account in typical cybersecurity countermeasures for computer networks.
    \item \textit{ISP-level layer} is organized at ISP (Internet Service Provider), which is responsible for detecting, filtering, and blocking verified domains of disinformation 2.0 (this is already done, e.g., for phishing emails, suspicious links, or blacklisted domains). In such a scenario, the ISP can be considered a  proxy between the users and the servers of the social network, located somewhere on the Internet.
    \item \textit{Device-level layer} is organized on the user's machine, typically within a browser or mobile apps, as this is how the user interacts with various websites and social networks. The security mechanisms deployed on this level should include automatic (e.g., AI-based) deepfake image or AI-generated text detection, in-browser news source verificationtion and 
 cross-verification of suspicious news at several trusted sources, and means to enforce user's responsible news sharing behavior (e.g., alerting the user when he/she tries to spread the news marked as potentially fake).
    \item \textit{User-level layer} is an essential part of the holistic approach to addressing disinformation 2.0, incorporating all manual actions that can be performed by users. For instance, this includes engaging with prebunking \cite{Roozenbeek2020} to raise the ability of the users to detect disinformation. Furthermore, given the rise and value of citizen journalism, it is important to empower users to perform disinformation detection using technical means that are commonly accessible to professional journalists in newsrooms. We consider that the long-standing goal of educating the users about disinformation by empowering them is the best way to ensure their resilience in the long-term.
\end{itemize}

Note that in order to be effective, all security mechanisms and user actions need to be applied in tandem. Moreover, employed mechanisms should be as diverse as possible, i.e., they preferably should base their detection approaches on different aspects of disinformation. 
It is also worth emphasizing that currently, not all of the above-mentioned methods to fight disinformation are in use (e.g., an automatic AI-based cross-verification of the news at different sources). Moreover, in Fig. \ref{fig:fakeproposed}, arrows between the layers indicate that each layer can transfer certain information to the other layer. For instance, disinformation detection mechanisms on social networks can tag a video or image as questionable when passing the news to the user's browser/app for further probing by the next layer, or it can filter it out and send down only a proper notification. On the other hand, if disinformation is discovered on the user's device level, 
then this information can be passed to the social network operator and displayed to the user. 

Some of the solutions in the proposed approach may be considered invasive from a user's privacy point of view, and protecting user data privacy is a critical concern in today's digital landscape. Fortunately, several effective solutions and strategies already exist that can be employed to fix this issue and follow the privacy-by-design principle, e.g., by incorporating schemes relying on differential privacy, secure data aggregation, homomorphic encryption, or data masking and tokenization.

In conclusion, we strongly believe that the advanced AI techniques, despite their benefits to society, greatly enable adversaries to achieve more sophisticated and effective disinformation 2.0 attacks. As such, adopting the lessons learned from cybersecurity research, novel countermeasures are needed, especially a holistic layered approach as discussed.

\begin{acks}
WM acknowledges the funding obtained from the EIG {CONCERT}-Japan call to the project Detection of disinformation on SocIal MedIa pLAtfoRms ``DISSIMILAR'' through grant EIG CONCERT-JAPAN/05/2021 (National Centre for Research and Development, Poland).
DL was in part supported by NSF awards \#1820609, \#2114824,
and \#2131144.
AV acknowledges the funding from ERC grant AVERITEC (GA 865958), and the EU H2020 grant MONITIO (GA 965576).
\end{acks}

\bibliographystyle{ACM-Reference-Format}
\bibliography{bibliography}

\end{document}